# Interfacially enhanced superconductivity in Fe(Te,Se)/Bi$_4$Te$_3$ heterostructures

*An-Hsi Chen[1], Qiangsheng Lu[1], Eitan Hershkovitz[2], Miguel L. Crespillo[3], Alessandro R. Mazza[1,4], Tyler Smith[1], T. Zac Ward[1], Gyula Eres[1], Shornam Gandhi[2], Meer Muhtasim Mahfuz[2], Vitalii Starchenko[5], Khalid Hattar[3], Joon Sue Lee[6], Honggyu Kim[2], Robert G. Moore[1*], Matthew Brahlek[1#]*

[1]Materials Science and Technology Division, Oak Ridge National Laboratory, Oak Ridge, TN, 37831, USA
[2]Department of Materials Science and Engineering, University of Florida, Gainesville, FL, USA
[3]Department Nuclear Engineering, The University of Tennessee, Knoxville, TN, 37996, USA
[4]Materials Science and Technology Division, Los Alamos National Laboratory, Los Alamos, NM, 87545, USA
[5]Chemical Sciences Division, Oak Ridge National Laboratory, Oak Ridge, TN, 37831, USA
[6]Department of Physics and Astronomy, The University of Tennessee, Knoxville, TN, 37996, USA
Correspondence should be addressed to *moorerg@ornl.gov, #brahlekm@ornl.gov

**Abstract**: Realizing topological superconductivity by integrating high-transition-temperature ($T_C$) superconductors with topological insulators can open new paths for quantum computing applications. Here, we report a new approach for increasing the superconducting transition temperature ($T_C^{onset}$) by interfacing the unconventional superconductor Fe(Te,Se) with the topological insulator Bi-Te system in the low-Se doping regime, near where superconductivity vanishes in the bulk. The critical finding is that the $T_C^{onset}$ of Fe(Te,Se) increases from nominally non-superconducting to as high as 12.5 K when Bi$_2$Te$_3$ is replaced with the topological phase Bi$_4$Te$_3$. Interfacing Fe(Te,Se) with Bi$_4$Te$_3$ is also found to be critical for stabilizing superconductivity in monolayer films where $T_C^{onset}$ can be as high as 6 K. Measurements of the electronic and crystalline structure of the Bi$_4$Te$_3$ layer reveal that a large electron transfer, epitaxial strain, and novel chemical reduction processes are critical factors for the enhancement of superconductivity. This novel route for enhancing $T_C$ in an important epitaxial system provides new insight on the nature of interfacial superconductivity and a platform to identify and utilize new electronic phases.

1. **Introduction**

Harnessing and understanding superconductivity has been a long-sought goal since its discovery in mercury in 1911(see Ref. [1] and references therein). Recently, the discovery of new superconducting platforms that may host topological states that have been predicted and may enable applications for error-tolerant quantum computers[2,3]. Although there have been proposals for topological superconductivity in homogeneous bulk crystals, many routes have been proposed to achieve the same physics by joining superconductors and topological insulators (TIs) at high-quality epitaxial interfaces[4,5]. Moreover, epitaxy offers many routes to tune both the superconductor and the TI via doping, strain, or band structure modification, which are critical to overcome fundamental materials challenges[6]. Key examples include elemental superconductors grown on group III-V semiconductors, such as Pt/Al/InSb[7], Pb/InAs[8], Sn/InSb[9], or the iron-based superconductors on TIs, $FeTe_{1-x}Se_x$/$Bi_2Te_3$[10–13]. However, elemental superconductors have low $T_C$ (highest $T_C$ is Nb ≈ 9 K), few tuning parameters, and are difficult to integrate with chalcogenide topological materials[14], which together limit the ability to test theoretical predictions and realize new applications. These factors have motivated the search for unconventional chalcogenide superconductors with higher $T_C$, which are typically formed by chemical doping. Although chemical doping enables tuning the superconductor properties, understanding the emergent phenomena remains challenging because of the unusual physics made more difficult by the complex role defects play in the properties.

Although FeSe and FeTe share the same tetragonal crystal structure, they have very different properties: FeSe is a superconductor with $T_C$ = 8 K, while FeTe is insulating and antiferromagnetic with $T_N$ ≈ 70 K[15,16]. Interestingly, substitutional Se doping of FeTe can lead to a $T_C$ higher than that in the pure FeSe phase. Bulk samples of $FeTe_{1-x}Se_x$ (FTS) show onset of superconductivity near $x$ ≈ 0.10, and a maximum $T_C$ ≈ 14 K at $x$ = 0.50[17,18]. The superconductivity is highly pressure and strain dependent with $T_C$ showing enhancements from 8 K to near 30 K in FeSe under pressure of up to 4.15 GPa, and 15.9 K in exfoliated $FeTe_{0.5}Se_{0.5}$ flakes when compressive strain is applied[19,20]. The superconductivity in monolayer thin films of FeSe has been found to be enhanced when interfaced with oxide substrates[21,22]. Interestingly, for epitaxial growth, FTS has a square surface lattice while $Bi_2Te_3$ has a triangular surface lattice, but, despite this in-plane symmetry mismatch, FTS and $Bi_2Te_3$ are epitaxially compatible[11,23]. The early report of epitaxial $Bi_2Te_3$/FeTe thin films by He *et al*. in 2014 demonstrated that superconductivity was induced when 1-5 quintuple layers (QL, corresponds to a monolayer of $Bi_2Te_3$) of the topological insulator $Bi_2Te_3$ were grown on top of FeTe[23]. In 2018, Chen *et al*. reported that molecular beam epitaxy (MBE) grown $Bi_2Te_3$ thin films on $FeTe_{0.55}Se_{0.45}$ have a superconducting gap and anisotropic vortices in the TI layer using scanning tunneling microscopy (STM), which suggest possible topological superconductivity[10]. In 2021, X. Yao *et al*. reported the reverse structure, where FTS was grown epitaxially on $Bi_2Te_3$. Despite the structural mismatch, they attributed the successful epitaxial integration of these materials to a close lattice match along a single crystalline axis, and called this "hybrid epitaxy"[11]. Most interestingly, FTS/$Bi_2Te_3$ heterostructures exhibited superconductivity at low Se content ($x < 0.1$), which suggests a novel interaction between the topological layer and the Fe-chalcogenide superconductor. Motivated by investigating candidate Majorana bound states in the clean limit, Moore *et al*. explored the low Se doping regime and demonstrated that the relative band-structures of both FTS and $Bi_2Te_3$ can be systematically tuned at the synthesis step via reduction of the $Bi_2Te_3$ and through Se doping[13]. The ability to tune the properties via MBE synthesis enables confirming that the desired spin-momentum locking can be achieved for low Se content ($x = 0.1$) in FTS as thin as a monolayer using spin and angle-resolved photoemission spectroscopy (ARPES). All together these early works highlight the intricate chemical and structural properties at interfaces with TIs with dissimilar materials, which can combine to affect the properties of both the TI and functional material[14,24–27]. As such, the link between material properties and the Se doping level in FTS on the emergent superconductivity outside of the bulk phase diagram in the low Se doping regime is not comprehensively understood.

Here, we find that the superconductivity in the FTS/Bi-Te system (herein we will refer to several phases within the Bi-Te family, $Bi_2Te_3$, $Bi_1Te_1$ and $Bi_4Te_3$) can be enhanced by tailoring the properties of the TI layer by MBE growth in a way that is not accessible in the bulk phase diagram. By utilizing a combination of structural, compositional, and electronic probes, it is specifically found that in the low Se doping limit $T_C$ is increased to 12.5 K when the $Bi_2Te_3$ is replaced with the topological material $Bi_4Te_3$. In contrast, optimally grown FTS/Bi$_2$Te$_3$ is found to closely follow the bulk property where the superconductivity vanishes with reducing Se content below $x \approx 0.15$. More importantly, superconductivity is found to extend down to the monolayer limit where the transition to zero resistance is enhanced only at the interface between FTS and Bi$_4$Te$_3$, which implies an interfacial origin for the enhancement. *In situ* ARPES measurements indicate that the Bi$_4$Te$_3$ phase is topological, in agreement with previous studies[28,29], as well as exhibiting a significantly larger Fermi surface than Bi$_2$Te$_3$. This enhanced density of states at the interface likely provides a large electron reservoir that acts together with strain, and chemical reduction across this interface to drive the enhanced superconductivity. This work shows a new route to simultaneously tune both superconductivity and topology and highlights novel physics that is accessible only through thin film synthesis routes.

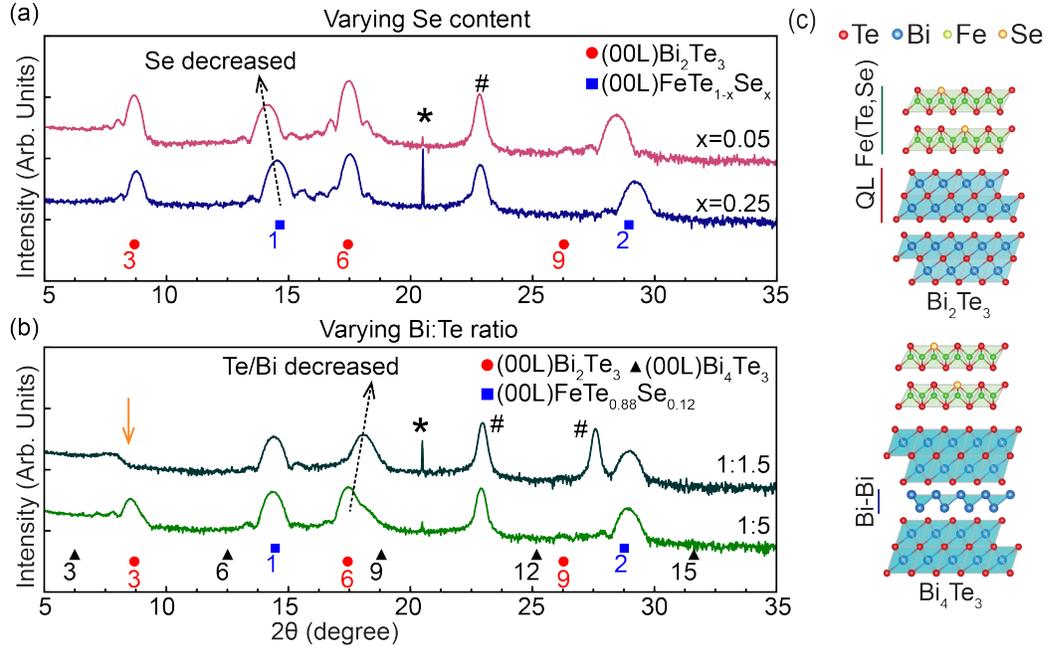

**Figure 1.** (a) XRD $2\theta$-$\theta$ measurement of FTS/Bi$_2$Te$_3$ as a function of Se content ($x$), and (b) with $x$ fixed at 0.12 versus Bi:Te flux ratio. The black dashed arrow in (a) marks the shift of the FTS (001) peak and in (b) the shift of (006) Bi$_2$Te$_3$ peak, the yellow arrow in (b) marks the disappearance of the Bi$_2$Te$_3$ (003) peak. In both (a) and (b) the blue squares mark {001} FTS reflections, red circles mark {003} Bi$_2$Te$_3$ reflections, and the black triangles mark {003} Bi$_4$Te$_3$ reflections. The # symbol indicates Te reflections of the capping layer and the * symbol marks the Al$_2$O$_3$ substrate reflections. (c) Schematic Bi$_2$Te$_3$ and Bi$_4$Te$_3$ structures.

## 2. Results

To understand the interactions between the FTS and the Bi-Te layer, two series of samples were studied. The first series focused on the emergence of superconductivity as a function of the Se concentration in the range $x = 0.00\text{-}0.50$ for FTS grown on stochiometric $Bi_2Te_3$ that was optimized for structural and electronic quality. The second series of samples consisted of FTS with $x$ fixed at $x = 0.12$ grown on the buffer Bi-Te layers ranging from $Bi_2Te_3$ to $Bi_4Te_3$. The stoichiometry of the Bi-Te layer was controlled by varying the (Bi:Te) flux ratio supplied to the growing sample from greater than (1:5) to as low as (1:0.5) during the growth. For the Bi-Te layer grown with (Bi:Te=1:5) flux ratio the phase is $Bi_2Te_3$ and (Bi:Te=1:1.5) is $Bi_4Te_3$ as shown later in the manuscript. The detailed growth parameters are given in the Supporting Information. Following growth all samples were capped with Te to prevent contamination and possible extrinsic routes that convolute the appearance of superconductivity such as oxidation or hydration[30–32] (see **Figure S8** which shows that the properties are stable for many months).

We first discuss the structural properties of a series of samples while varying the Se content in the FTS layer in the range $x = 0.05\text{-}0.50$. X-ray diffraction (XRD) $2\theta\text{-}\theta$ scans are shown in **Figure 1a** where the {001} set of diffraction peaks for FTS, $Bi_2Te_3$ and $Al_2O_3$ substrate are observed. Note that Te was deposited at room temperature and is textured with a dominant orientation of {100} with additional orientations observed occasionally. The peak positions and corresponding lattice parameters of $Bi_2Te_3$ and FTS are consistent with those of bulk materials. Specifically, the $Bi_2Te_3$ peak position gives the out-of-plane lattice parameter $c_{Bi2Te3} \approx 3.04$ nm (QL$\approx 1.01$ nm). According to Vegard's law and assuming no strain[33] the FTS out-of-plane lattice parameter changes linearly with composition and is given by $c_{FTS}(x) \approx c_{FeSe}(x) + c_{FeTe}(1\text{-}x)$, which is manifested by a peak shift toward higher $2\theta$ with increasing Se concentration. The lattice constant $c_{FeTe} \approx 0.63$ nm is found to be consistent to the bulk value of $\sim 0.63$ nm, which is within the error range estimated for the values reported here. The prominent Pendellösung fringes observed about the main peaks of both $Bi_2Te_3$ and FTS indicate high crystalline quality and near atomically flat interfaces[34]. These data demonstrate that this growth recipe produces the highest crystalline quality films with sharp interfaces.

The high sensitivity to charge and strain may be used to modify the superconducting character of FTS by changing the carrier concentration and the structure through tailoring the properties of the Bi-Te layer. The $A_2X_3$ tetradymite TIs $Bi_2Te_3$, $Bi_2Se_3$ and $Sb_2Te_3$ are characterized by 5 atom QLs, as shown at the top of **Figure 1c**. This motif permits structural and charge tuning through reduction of the chalcogenide,

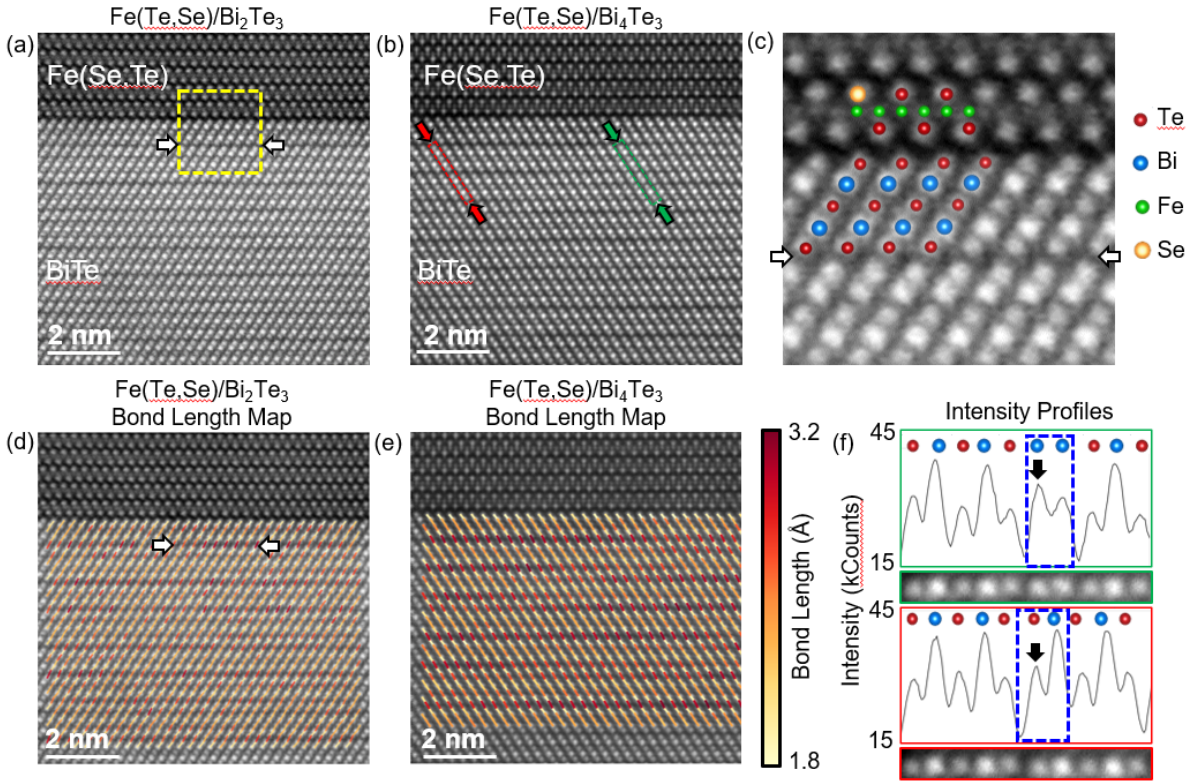

**Figure 2.** (a-b) HAADF-STEM images of the FTS/BiTe interface for the films grown with Bi:Te=1:5 (a, $Bi_2Te_3$) and Bi:Te=1:1.5 (b, $Bi_4Te_3$). These were imaged along the [100] zone axis of the FTS/Bi-Te, respectively. (c) Magnified image of the yellow boxed region in (a) with atomic schematic overlayed. (d-e) Bond length maps of the Bi-Te films measured from (a) and (b), respectively. (f) Intensity profiles of the red and green boxed regions in (b). The dashed blue box corresponds to the same row of atoms in both intensity profiles illustrating the change in intensity for the atomic columns indicative of the heavier Bi replacing the lighter Te atom. See **Figure 1** for XRD of these films.

Se or Te[35,36]. As shown at the bottom of **Figure 1c**, the central difference among the structures is the additional bismuth bilayer that is intercalated between QLs ($Bi_4Te_3$) or alternating QLs ($Bi_1Te_1$)[28]. The phase control among these various phases can be easily achieved by varying the (Bi:Te) flux ratio during the growth from Te-rich (1:5) to Te-deficient (1:1.5) conditions or by postgrowth annealing[13,37]. The XRD $2\theta$-$\theta$ scans in **Figure 1b** show subtle changes relative to $Bi_2Te_3$ that suggest the formation of a bismuth bilayer. Specifically, the (003) $Bi_2Te_3$ peak clearly vanishes and the (006) peak shifts slightly toward higher $2\theta$, becoming the (009) $Bi_4Te_3$. Additionally, at higher $2\theta$ (**Figure S7**), additional peaks shift to lower $2\theta$ which indicates the phase transition instead of strain or defect accumulation. The $Bi_4Te_3$ peak appears wider and more asymmetric in XRD than optimally grown $Bi_2Te_3$, which likely stems from possible crystalline disorder associated with the bismuth bilayer formation. Moreover, the difference of (009) $Bi_4Te_3$ in $2\theta$ between XRD data and bulk data is probably due to the inhomogeneous distribution of bismuth bilayers and different types and densities of defects in bulk crystals compared to epitaxial films[35]. However, the FTS remains unchanged retaining clear Pendellösung fringes that correspond to highly uniform surfaces. Rutherford backscattering spectroscopy (RBS) used for measuring the chemical composition also supports the crystalline phase change from $Bi_2Te_3$ to $Bi_4Te_3$ with decreasing (Bi:Te) flux ratio, as shown in **Table 1**. The lattice parameter of FTS is close to the bulk value for $x \approx 0.12$. However, with decreasing Te content in the buffer layer, there is a slight shift towards higher $2\theta$, with the lattice parameter shown in **Figure S4**. Interestingly, this reduction of the FTS interplanar distance can be caused by in-plane tensile strain or perhaps slight anion diffusion from FTS to the $Bi_4Te_3$ layer. For example, Te readily diffuses across this interface, which is shown in **Figures S5** and **S6** of the Supporting Information, where the $Bi_2Te_3$ is reduced when FeTe was grown in a Te-deficient regime.

Atomic resolution high-angle annular dark-field (HAADF) images acquired by scanning transmission electron microscopy (STEM) along the [100] zone axis of the FTS/Bi-Te interface are shown in **Figure 2a** for two samples grown with (Bi:Te=1:5) corresponding to $Bi_2Te_3$ and in **Figure 2b** for (1:1.5) ratio corresponding to $Bi_4Te_3$. **Figure 2c** shows a magnified image of the interface marked by a yellow box in **Figure 2a**. These images reveal critical structural features of these interfaces. First, the high magnification HAADF-STEM image in **Figure 2c** demonstrates a chemically abrupt and coherent interface between the FTS and Bi-Te films, which appears to be always terminated by a QL for $Bi_4Te_3$. This suggests that regardless of the flux ratio used during the growth of the Bi-Te layer, there is sufficient Te in the system to turn the Bi-excess into a QL, which likely implies that the FTS interface layer is reduced. Second, HAADF-STEM images provide snapshots of the local microstructure. Specifically, these data give insight into the formation of the bismuth bilayer. The white arrows in **Figures 2a**, **2c**, and **2d** correspond to the same location highlighting the dark streak that is the result of a large Te-Te bond distance between two QLs. **Figure 2d** for Bi$_2$Te$_3$ and **Figure 2e** for Bi$_4$Te$_3$ show the bond distances between neighboring atomic columns overlaid on the HAADF-STEM images of **Figures 2a** and **2b**, respectively. Here, the color of the bonds corresponds to their length with red representing longer bonds and yellow representing shorter bonds. Color coding the bond lengths is used to illustrate clearly the QL-QL gaps which have a larger projected spacing than the gap between the bismuth bilayer and the QL. This agrees with the bulk structure where the Te-Te van der Waals bond length between the QLs is the largest at 3.58 Å while the bond lengths of the Bi-bismuth bilayer and the Bi-Te bond within the Bi$_2$Te$_3$ QL are 3.29 Å and 3.12 Å, respectively. **Figure 2f** shows two intensity profiles from the sample shown in **Figure 2b** illustrating the difference between the nominal QL (lower red box) and an atomic layering (upper green box) that includes the QL and a bismuth bilayer. The chemical sensitivity of HAADF-STEM imaging shows an increase in the image intensity of the atomic column where the heavier Bi atoms supplant the lighter Te atoms indicated by the black arrows. This atomic column is closer to the preceding QL, resulting in a short bond length. The reduction of Bi$_2$Te$_3$ to Bi$_4$Te$_3$ seen across the sample is analyzed in **Figure S1** where a robust statistical comparison between the two samples is made and we can see that the nominal $Bi_2Te_3$ sample is quite uniform, while the sample with (Bi:Te=1:1.5) shows some Bi richness indicative of a larger fraction of $Bi_4Te_3$. This agrees well with the structure reported by X. Yao *et al.* showing partial formation of the bismuth bilayer[11], as well as the macroscale XRD data shown in **Figure 1b** on the same samples. Finally, as shown in **Figure S3**, novel structures where the top of the QL unit merges seamlessly with the FTS monolayer are occasionally found, as previously reported in STM experiments[38].

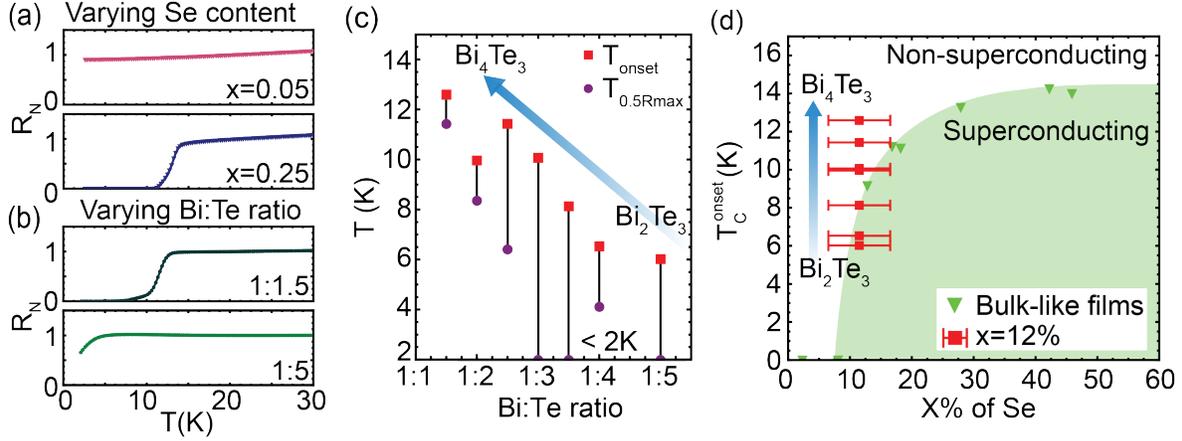

**Figure 3.** (a) Normalized resistance versus temperature for FTS/Bi$_2$Te$_3$ versus Se content ($x$) and (b) with $x$ fixed at 0.12 versus (Bi:Te) flux ratio. (c) $T_C$ versus (Bi:Te) ratio of Bi$_2$Te$_3$ layer. The red squares represent the superconducting onset temperature $T_{onset}$, and the purple circles represent the half-resistance temperature $T_{0.5Rmax}$. (d) The summarized $T_{onset}$ versus Se content $x$ for both Se content series (green inverted triangle) and (Bi:Te) ratio series (red squares). The green color marks the superconducting region, and the white color marks the non-superconducting region for bulk.

The measurements of resistance versus temperature in **Figure 3** indicate the changes of superconductivity for bulk-like films as well as for the Te-reduced films. The resistance is normalized by the value at 15 K, which is well above the onset of superconductivity, $T_C^{onset}$ (temperature at which the resistance is 90% of the resistance at 15 K). The data for FTS on $Bi_2Te_3$ are shown in **Figure 3a** where the drop to zero resistance indicates the emergence of superconductivity for $x = 0.25$ compared to the sample with $x = 0.05$ that is clearly not superconducting. This is consistent with the trends in bulk FTS as a function of Se content with superconductivity vanishing near 10-20% Se concentrations. Note that the estimated stoichiometry in bulk crystals may be systematically inaccurate because of uncertainties in the calibration method, primarily energy dispersive spectroscopy as well as the final (Te:Se) ratio relative to the starting composition[17,39]. In contrast to the bulk-like films, the resistance versus temperature curve in **Figure 3b** shows that superconductivity is clearly enhanced in samples with a $Bi_4Te_3$ buffer layer. This is illustrated in **Figure 3c** where $T_C$ clearly rises with decreasing Te flux during the Bi-Te layer growth. Here, the superconducting transition temperature is indicated by both $T_C^{onset}$ (red squares) and $T_{0.5Rmax}$ (purple circles, where the resistance is 50% of the resistance at 15 K). For samples with (1:3), (1:3.5), and (1:5) (Bi:Te) ratio the $T_{0.5Rmax}$ are limited at 2 K that is the lower limit in the cryostat used. Interestingly, the sharpness of the onset increases for the lowest (Bi:Te) ratio. The sharper onset may be related to the superconducting order being strongest when $Bi_4Te_3$ has minimal crystalline disorder. The data for FTS on both $Bi_2Te_3$ and $Bi_4Te_3$ are summarized as $T_C$ versus $x$ in the phase diagram in **Figure 3d**. Here, the optimized samples with varying Se content are shown as green triangles and the Te-reduced series are shown as red squares. Note that the uncertainty in the Se content of all samples determined by RBS is around 5%. These data clearly show that with fixed Se concentration in the low Se regime, the FTS on $Bi_4Te_3$ has much higher superconducting temperatures than the FTS on $Bi_2Te_3$, which highlights a strong coupling that is unique to this interface.

One of the most significant results of this work is that the enhancement of superconductivity at the $Bi_4Te_3$ interface was found to be critical to stabilize emergent superconductivity in monolayer FTS. To show this stabilization, an FTS ($x = 0.25$) film which was slightly thicker than a monolayer was grown on a 15 nm $Bi_2Te_3$ film. The $Bi_4Te_3$ phase was achieved by annealing $Bi_2Te_3$ at 225°C without Te, as reported previously[13,37]. The XRD data in **Figure 4a** show that the (006) peak of the $Bi_2Te_3$ shifts toward the (009) $Bi_4Te_3$ and the (003) $Bi_2Te_3$ nearly vanishes, again indicating the formation of the $Bi_4Te_3$ phase. The control sample is $Bi_2Te_3$. Both samples show Pendellösung fringes indicating near atomically sharp interfaces. Also, the FTS peaks are not resolved since they are monolayers. Comparison of transport measurements, shown in **Figure 4b**, confirm that the FTS grown on $Bi_4Te_3$ has a higher $T_C^{onset}$ than the sample on $Bi_2Te_3$, reaching nearly zero resistance at 2.5 K. These data confirm that the superconducting properties of FTS are strongly localized at the interface.

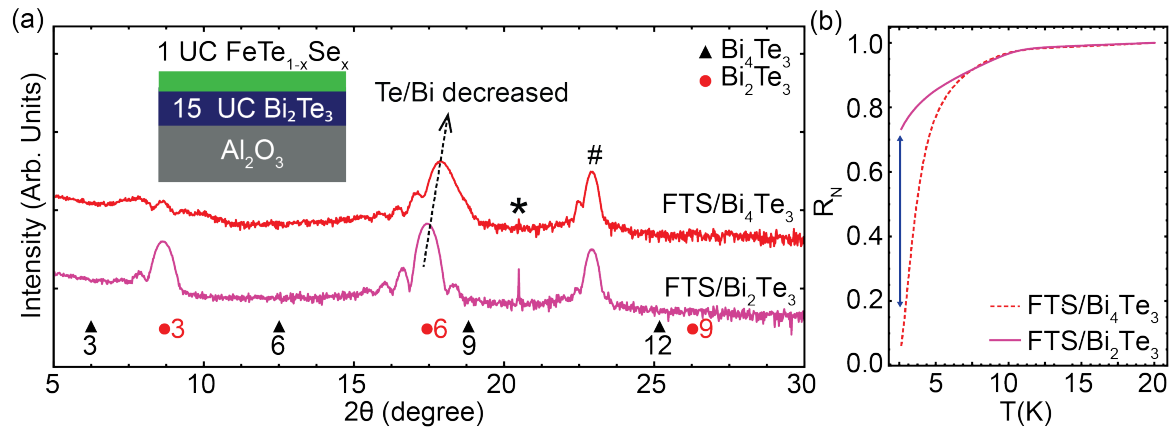

**Figure 4.** (a) XRD $2\theta$-$\theta$ measurement of monolayer FTS grown on $Bi_2Te_3$ or $Bi_4Te_3$. The red circles mark {003} $Bi_2Te_3$ reflection, black triangles mark {003} $Bi_4Te_3$ reflections, the hash symbol indicates Te reflections from the capping layer and the asterisk symbol marks the $Al_2O_3$ substrate reflections. The black dashed arrow highlights the (006) $Bi_2Te_3$ peak shift. (b) The plot of normalized resistance versus temperature, FTS/$Bi_2Te_3$ in pink sold curve and FTS/$Bi_4Te_3$ in red dashed curve. The blue arrow marks the resistance difference at low temperature.

The electronic structure of Te-reduced $Bi_2Te_3$ was characterized by *in situ* ARPES measurements that were performed on a series of Bi-Te films grown with different (Bi:Te) ratios in a range from (1:20) to (1:0.5). The top row of **Figure 5a** shows the ARPES spectra at the Fermi surface as determined by integrating a 40 meV window centered at the Fermi energy, and the bottom row shows the band structure cuts along the $\Gamma$-$K$ direction. The white dashed line in the top row marks the measurement position of the electronic dispersion shown in the bottom row. There are two significant changes observed with decreasing Te composition corresponding to the evolution of the band structure, and the Fermi surface size/density of states. We observe two different surface states, SS1 (green dashed line) and SS2 (blue dashed line). In the (1:20) sample, the band structure SS1 has a clear Dirac cone at $E$ = -0.2 eV at the $\Gamma$ point, and a hexagonal Fermi surface, consistent with previous $Bi_2Te_3$ results[40]. With decreasing Te flux, the second surface state SS2 appears, where the Fermi surface enlarges into a star shape, in agreement with the recent ARPES study on bulk $Bi_4Te_3$[29]. The new surface state (SS2) could change the net properties of the interface by hybridizing with FTS states forming spin-momentum locked electron band structures similar to previous studies of the FTS/$Bi_2Te_3$ interface where the TI was just annealed[13]. It is important to emphasize that the band structures are clearly resolved while the material structurally evolves between the different motifs. A deeper understanding of the spin-polarized properties is needed to understand the viability of this as a topological superconducting candidate system. The second major change is that the valence band shifts downward with decreasing (Bi:Te) ratio, marked with red arrows. This band shift indicates that strong electron-doping occurs during the Te reduction process, explaining the large increase in the density of states at the Fermi level in the Bi-Te layer. This additional source of electrons will be electrostatically transferred into FTS, which likely has a strong effect on driving up the $T_C$ in low-Se-doped FTS.

3. **Discussion**

We have systematically explored the structure, transport, and electronic properties of FTS/$Bi_2Te_3$ heterostructures and found a strong enhancement of the superconducting transition temperature when the $Bi_2Te_3$ layer is reduced to $Bi_4Te_3$ that is critical for stabilizing superconductivity in the monolayer limit. In bulk and thin films superconductivity in FTS is known to be strongly affected by charge, strain[42], and at interfaces with different materials such as $SrTiO_3$[21,22] where a novel phonon enhancement increases the transition temperature. Our data show that many of these factors may be critical for the manifestation of superconductivity. First, as shown in **Figure 5b**, a large charge reservoir forms when the $Bi_2Te_3$ is reduced to $Bi_4Te_3$. Since the charge in the FTS is nominally unchanged the relatively higher Fermi level in the $Bi_4Te_3$ will force the charge to spill into the FTS layer, thus causing dramatic n-type doping. Specifically, FTS/$Bi_2Te_3$ exhibits a sheet carrier density of $3.5\times10^{14}$ cm$^{-2}$ whereas in FTS/$Bi_4Te_3$ is increased to $3.3\times10^{15}$ cm$^{-2}$. Hall effect measurements sum the carriers across the entire structure, so it remains unknown what the actual density is in the FTS. It is interesting to note that $T_C$ in bulk FTS typically increases due to strong p-type doping that occurs as a result of mixing Se into FeTe[43,44]. However, $Bi_4Te_3$ transfers electrons, which is the opposite compared to the bulk.

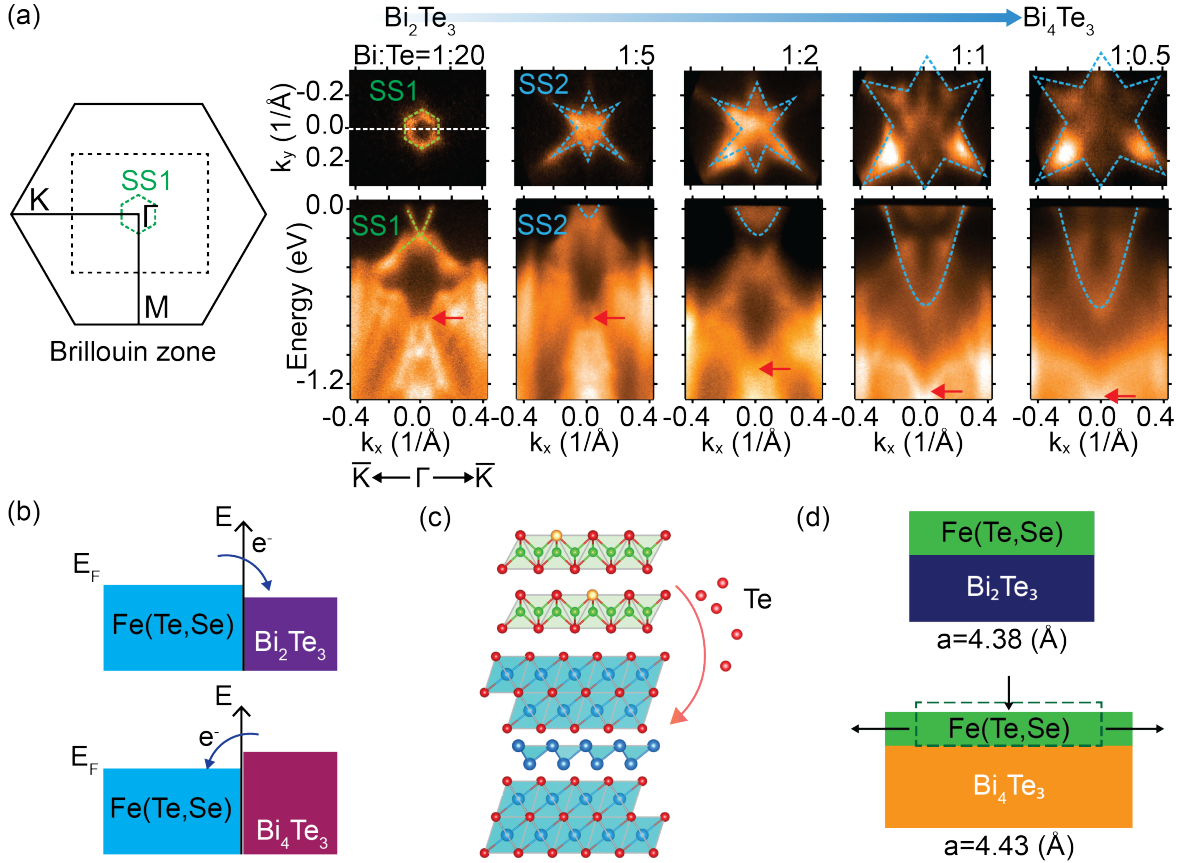

**Figure 5**. (a) In-situ ARPES measurements on $Bi_2Te_3$ layer with different Bi:Te ratios. The schematic Brillouin Zone at left-hand-side shows the size (dashed square) of ARPES measurements along high symmetry points. The green dashed hexagon represents the size of surface state 1(SS1). The top row shows the Fermi surface, momentum $k_y$ versus $k_x$, and the bottom row shows the band structure along the line marks out in the top row. SS1 (green dashed line) and SS2 (blue dashed line) indicate two different kinds of electronic surface states. The red arrows indicate the energy position of bulk band structure. (b-d) Schematics of the possible mechanism that drive Te-vacancy enhanced superconductivity. (b) Higher relative Fermi level in $Bi_4Te_3$ may drive strong charge transfer, (c) Te/Se may diffuse from the FTS to $Bi_2Te_3$ attaining a different Fe:(Te+Se) ratio that what is different in the bulk, and (d) strain-induced in FTS.

It was found that for fixed $x$ the c-axis lattice parameter of FTS was slightly reduced when grown on $Bi_4Te_3$ (**Figure 1b** and **Figures S4-S6**). This points to either strain or a novel chemical reduction process. Regarding strain, as the bulk in-plane lattice parameter of $Bi_2Te_3$ is 4.42 Å[45] and $Bi_4Te_3$ is 4.47 Å[46], there may be residual in-plane tensile strain that reduces the out-of-plane lattice parameter of FTS, as schematically shown in **Figure 5d**. FTS is known to be highly sensitive to strain, where the transition temperature can be increased from 12 K to 21 K with the application of ~2% compressive strain[42]. Regarding the chemical reduction, in bulk FTS crystals naturally grow with excess Fe. Here, it is plausible that the cation/anion ratio is altered for the $Bi_4Te_3$ system since the synthesis temperature and conditions are significantly different compared to both bulk crystals and even the optimally grown FTS/$Bi_2Te_3$. As shown in **Figure 5c**, this may imply that there is a large chemical potential ($\mu_{TE}$) that drives the chalcogenide from the FTS into the $Bi_4Te_3$, which may result in a Fe/(Te+Se) ratio that is significantly different from what is achieved in bulk. This may also provide additional insight into how the superconductivity emerges in the inverted structure, $Bi_2Te_3$ on FeTe[47].

Finally, the data presented here on monolayer films indicate that the interfacial properties have a strong effect on superconductivity. Therefore, it is important to understand the role that the termination of the $Bi_4Te_3$ layer plays. Specifically, $Bi_4Te_3$ can either be terminated with the bismuth bilayer or with a QL. The HAADF-STEM data in **Figure 2** show predominantly a QL termination. The enthalpy of formation is lower for $Bi_2Te_3$ than for $Bi_4Te_3$[48]. This indicates that there may be a strong driving force for diffusion of Te from FTS to $Bi_4Te_3$ relative to $Bi_2Te_3$, which is consistent with the observations that Te migrates across this interface. This indicates that the FTS at the interface is likely highly reduced, yet remains intact structurally. Further investigations are needed to directly observe the role of the large change in chemical potential of FTS grown on $Bi_2Te_3$ with different (Bi:Te) ratios. Overall, the role this plays in superconductivity is currently not known. Therefore, it is important to explore possible interfaces between the FTS layer with bismuth bilayer termination, and whether this can be stabilized with the natural drive to pull Te from the FTS layer. Future studies should include systematic experimental efforts to study stability of intentionally terminated surfaces with Bi and QL when interfaced with FTS, and theory investigations to understand the kinetics and thermodynamics as well as how the interfacial properties affect superconductivity.

## 4. Conclusion

Our data have shown a new route to increase $T_C$ in FTS by tailoring the properties of the underlying TI layer. Specifically, it is found that reducing the layer towards $Bi_4Te_3$ increases the transition temperature for samples with low Se doping. Moreover, this has proven critical for stabilizing $T_C$ in the monolayer limit, which clearly points to a strong interfacial effect. Although the detailed mechanism is currently not clear, the data indicate strong interplay between charge and strain, in controlling the properties of the superconducting interface. This highlights many interesting questions to better understand the origins for the enhanced superconductivity at this interface and how this new paradigm can be advanced. Specifically, deconvoluting the role of charge and strain can likely be done through, for example Se doping, $Bi_4Te_3$ termination as well as many other routes to tailor the interface. This work provides a clear set of design strategies that enable tunable topology and superconductivity, which are critical for new applications in quantum information science.

**Data Availability**

The data that support the findings of this study are available from the corresponding author upon reasonable request.

**Supporting Information**

Supporting Information includes

1. Molecular beam epitaxy growth
2. Experimental Methods
    a. X-ray diffraction
    b. Scanning transmission electron microscopy
    c. Transport
    d. Angle-resolved photoemission spectroscopy
    e. Rutherford backscattering spectroscopy
3. Additional structural data

As well as references [49–55]

**Acknowledgements**

This material was based on work supported by the U.S. Department of Energy, Office of Science, National Quantum Information Sciences Research Centers, Quantum Science Center (synthesis and spectroscopy), and from U.S. DOE, Office of Science, Basic Energy Sciences, Materials Science and Engineering Division (transport and structural characterization). ARM was funded by the NNSA's Laboratory Directed Research and Development Program at Los Alamos National Laboratory. Los Alamos National Laboratory, an affirmative action equal opportunity employer, is managed by Triad National Security, LLC for the U.S. Department of Energy's NNSA, under contract 89233218CNA000001. V.S. effort was sponsored by the Laboratory Directed Research and Development Program of Oak Ridge National Laboratory, managed by UT-Battelle, LLC, for the U.S. Department of Energy. We would like the thank Seongshik Oh, Xiong Yao, Michael McGuire, and Brain Sales for valuable discussions.

Received: ((will be filled in by the editorial staff))
Revised: ((will be filled in by the editorial staff))
Published online: ((will be filled in by the editorial staff))

References

[1] D. van Delft, P. Kes, *Phys. Today* **2010**, *63*, 38.
[2] A. Yu. Kitaev, *Ann. Phys.* **2003**, *303*, 2.
[3] C. Nayak, S. H. Simon, A. Stern, M. Freedman, S. Das Sarma, *Rev. Mod. Phys.* **2008**, *80*, 1083.
[4] L. Fu, C. L. Kane, *Phys. Rev. Lett.* **2008**, *100*, 096407.
[5] M. Sato, Y. Ando, *Rep. Prog. Phys.* **2017**, *80*, 076501.
[6] M. Brahlek, *Adv. Mater.* **2020**, *32*, 2005698.
[7] N. van Loo, G. P. Mazur, T. Dvir, G. Wang, R. C. Dekker, J.-Y. Wang, M. Lemang, C. Sfiligoj, A. Bordin, D. van Driel, G. Badawy, S. Gazibegovic, E. P. A. M. Bakkers, L. P. Kouwenhoven, *Nat. Commun.* **2023**, *14*, 3325.
[8] T. Kanne, M. Marnauza, D. Olsteins, D. J. Carrad, J. E. Sestoft, J. de Bruijckere, L. Zeng, E. Johnson, E. Olsson, K. Grove-Rasmussen, J. Nygård, *Nat. Nanotechnol.* **2021**, *16*, 776.
[9] M. Pendharkar, B. Zhang, H. Wu, A. Zarassi, P. Zhang, C. P. Dempsey, J. S. Lee, S. D. Harrington, G. Badawy, S. Gazibegovic, R. L. M. Op het Veld, M. Rossi, J. Jung, A.-H. Chen, M. A. Verheijen, M. Hocevar, E. P. A. M. Bakkers, C. J. Palmstrøm, S. M. Frolov, *Science* **2021**, *372*, 508.
[10] M. Chen, X. Chen, H. Yang, Z. Du, H.-H. Wen, *Sci. Adv.* **2018**, *4*, eaat1084.
[11] X. Yao, M. Brahlek, H. T. Yi, D. Jain, A. R. Mazza, M.-G. Han, S. Oh, *Nano Lett.* **2021**, *21*, 6518.
[12] S. Li, C. de la Cruz, Q. Huang, Y. Chen, J. W. Lynn, J. Hu, Y.-L. Huang, F.-C. Hsu, K.-W. Yeh, M.-K. Wu, P. Dai, *Phys. Rev. B* **2009**, *79*, 054503.
[13] R. G. Moore, Q. Lu, H. Jeon, X. Yao, T. Smith, Y.-Y. Pai, M. Chilcote, H. Miao, S. Okamoto, A.-P. Li, S. Oh, M. Brahlek, *Adv. Mater.* **2023**, *35*, 2210940.
[14] M. Brahlek, J. Lapano, J. S. Lee, *J. Appl. Phys.* **2020**, *128*, 210902.
[15] F.-C. Hsu, J.-Y. Luo, K.-W. Yeh, T.-K. Chen, T.-W. Huang, P. M. Wu, Y.-C. Lee, Y.-L. Huang, Y.-Y. Chu, D.-C. Yan, M.-K. Wu, *Proc. Natl. Acad. Sci.* **2008**, *105*, 14262.
[16] L. Kang, C. Ye, X. Zhao, X. Zhou, J. Hu, Q. Li, D. Liu, C. M. Das, J. Yang, D. Hu, J. Chen, X. Cao, Y. Zhang, M. Xu, J. Di, D. Tian, P. Song, G. Kutty, Q. Zeng, Q. Fu, Y. Deng, J. Zhou, A. Ariando, F. Miao, G. Hong, Y. Huang, S. J. Pennycook, K.-T. Yong, W. Ji, X. Renshaw Wang, Z. Liu, *Nat. Commun.* **2020**, *11*, 3729.
[17] M. H. Fang, H. M. Pham, B. Qian, T. J. Liu, E. K. Vehstedt, Y. Liu, L. Spinu, Z. Q. Mao, *Phys. Rev. B* **2008**, *78*, 224503.
[18] B. C. Sales, A. S. Sefat, M. A. McGuire, R. Y. Jin, D. Mandrus, Y. Mozharivskyj, *Phys. Rev. B* **2009**, *79*, 094521.
[19] S. Masaki, H. Kotegawa, Y. Hara, H. Tou, K. Murata, Y. Mizuguchi, Y. Takano, *J Phys Soc Jpn* **2009**, *78*.
[20] W. Wang, J. Li, J. Yang, C. Gu, X. Chen, Z. Zhang, X. Zhu, W. Lu, H.-B. Wang, P.-H. Wu, Z. Yang, M. Tian, Y. Zhang, V. V. Moshchalkov, *Appl. Phys. Lett.* **2014**, *105*, 232602.
[21] W. Qing-Yan, L. Zhi, Z. Wen-Hao, Z. Zuo-Cheng, Z. Jin-Song, L. Wei, D. Hao, O. Yun-Bo, D. Peng, C. Kai, W. Jing, S. Can-Li, H. Ke, J. Jin-Feng, M. Xu-Cun, X. Qi-Kun, *Chin. Phys. Lett.* **2012**, *29*, 037402.
[22] J. J. Lee, F. T. Schmitt, R. G. Moore, S. Johnston, Y.-T. Cui, W. Li, M. Yi, Z. K. Liu, M. Hashimoto, Y. Zhang, D. H. Lu, T. P. Devereaux, D.-H. Lee, Z.-X. Shen, *Nature* **2014**, *515*, 245.

[23] Q. L. He, H. Liu, M. He, Y. H. Lai, H. He, G. Wang, K. T. Law, R. Lortz, J. Wang, I. K. Sou, *Nat. Commun.* **2014**, *5*, 4247.
[24] L. A. Walsh, C. M. Smyth, A. T. Barton, Q. Wang, Z. Che, R. Yue, J. Kim, M. J. Kim, R. M. Wallace, C. L. Hinkle, *J. Phys. Chem. C* **2017**, *121*, 23551.
[25] E. Longo, C. Wiemer, R. Cecchini, M. Longo, A. Lamperti, A. Khanas, A. Zenkevich, M. Fanciulli, R. Mantovan, *J. Magn. Magn. Mater.* **2019**, *474*, 632.
[26] I. Vobornik, G. Panaccione, J. Fujii, Z.-H. Zhu, F. Offi, B. R. Salles, F. Borgatti, P. Torelli, J. P. Rueff, D. Ceolin, A. Artioli, M. Unnikrishnan, G. Levy, M. Marangolo, M. Eddrief, D. Krizmancic, H. Ji, A. Damascelli, G. van der Laan, R. G. Egdell, R. J. Cava, *J. Phys. Chem. C* **2014**, *118*, 12333.
[27] D. Flötotto, Y. Ota, Y. Bai, C. Zhang, K. Okazaki, A. Tsuzuki, T. Hashimoto, J. N. Eckstein, S. Shin, T.-C. Chiang, *Sci. Adv.* **2018**, *4*, eaar7214.
[28] T. Chagas, G. A. S. Ribeiro, P. H. R. Gonçalves, L. Calil, W. S. Silva, Â. Malachias, M. S. C. Mazzoni, R. Magalhães-Paniago, *Electron. Struct.* **2020**, *2*, 015002.
[29] T. Chagas, O. A. Ashour, G. A. S. Ribeiro, W. S. Silva, Z. Li, S. G. Louie, R. Magalhães-Paniago, Y. Petroff, *Phys. Rev. B* **2022**, *105*, L081409.
[30] K. Deguchi, T. Okuda, H. Hara, S. Demura, T. Watanabe, H. Okazaki, M. Fujioka, S. J. Denholme, T. Ozaki, T. Yamaguchi, H. Takeya, F. Saito, M. Hisamoto, Y. Takano, *Phys. C Supercond.* **2013**, *487*, 16.
[31] H. Hiramatsu, T. Katase, T. Kamiya, M. Hirano, H. Hosono, *Phys. Rev. B* **2009**, *80*, 052501.
[32] J. Hu, G. C. Wang, B. Qian, Z. Q. Mao, *Supercond Sci Technol* **2012**.
[33] M. Brahlek, L. Zhang, H.-T. Zhang, J. Lapano, L. R. Dedon, L. W. Martin, R. Engel-Herbert, *Appl. Phys. Lett.* **2016**, *109*, 101903.
[34] A. R. Lang, M. Zhen-Hong, *Proc. R. Soc. Lond. Ser. Math. Phys. Sci.* **1979**, *368*, 313.
[35] G. Springholz, S. Wimmer, H. Groiss, M. Albu, F. Hofer, O. Caha, D. Kriegner, J. Stangl, G. Bauer, V. Holý, *Phys. Rev. Mater.* **2018**, *2*, 054202.
[36] O. Caha, A. Dubroka, J. Humlíček, V. Holý, H. Steiner, M. Ul-Hassan, J. Sánchez-Barriga, O. Rader, T. N. Stanislavchuk, A. A. Sirenko, G. Bauer, G. Springholz, *Cryst. Growth Des.* **2013**, *13*, 3365.
[37] S. Kusaka, T. T. Sasaki, K. Sumida, S. Ichinokura, S. Ideta, K. Tanaka, K. Hono, T. Hirahara, *Appl. Phys. Lett.* **2022**, *120*, 173102.
[38] G. Chen, A. Aishwarya, M. R. Hirsbrunner, J. O. Rodriguez, L. Jiao, L. Dong, N. Mason, D. Van Harlingen, J. Harter, S. D. Wilson, T. L. Hughes, V. Madhavan, *Npj Quantum Mater.* **2022**, *7*, 1.
[39] T. Noji, T. Suzuki, H. Abe, T. Adachi, M. Kato, Y. Koike, *J. Phys. Soc. Jpn.* **2010**, *79*, 084711.
[40] Y. L. Chen, J. G. Analytis, J.-H. Chu, Z. K. Liu, S.-K. Mo, X. L. Qi, H. J. Zhang, D. H. Lu, X. Dai, Z. Fang, S. C. Zhang, I. R. Fisher, Z. Hussain, Z.-X. Shen, *Science* **2009**, *325*, 178.
[41] Q. Li, W. Si, I. K. Dimitrov, *Rep. Prog. Phys.* **2011**, *74*, 124510.
[42] E. Bellingeri, I. Pallecchi, R. Buzio, A. Gerbi, D. Marrè, M. R. Cimberle, M. Tropeano, M. Putti, A. Palenzona, C. Ferdeghini, *Appl. Phys. Lett.* **2010**, *96*, 102512.
[43] D. J. Gawryluk, J. Fink-Finowicki, A. Wiśniewski, R. Puźniak, V. Domukhovski, R. Diduszko, M. Kozłowski, M. Berkowski, *Supercond. Sci. Technol.* **2011**, *24*, 065011.
[44] Q. Ma, Q. Gao, W. Shan, X. Li, H. Li, Z. Ma, *Vacuum* **2022**, *195*, 110661.

[45] *Materials Data on Bi2Te3 by Materials Project*, Lawrence Berkeley National Lab. (LBNL), Berkeley, CA (United States). LBNL Materials Project, **2020**.
[46] *Materials Data on Bi4Te3 by Materials Project*, Lawrence Berkeley National Lab. (LBNL), Berkeley, CA (United States). LBNL Materials Project, **2020**.
[47] Q. L. He, H. Liu, M. He, Y. H. Lai, H. He, G. Wang, K. T. Law, R. Lortz, J. Wang, I. K. Sou, *Nat. Commun.* **2014**, *5*, 4247.
[48] W. Gierlotka, *Calphad* **2018**, *63*, 6.
[49] G. A. Norton, R. E. Daniel, R. L. Loger, J. B. Schroeder, *Nucl. Instrum. Methods Phys. Res. Sect. B Beam Interact. Mater. At.* **1989**, *37–38*, 403.
[50] J. B. Schroeder, C. W. Howell, G. A. Norton, *Nucl. Instrum. Methods Phys. Res. Sect. B Beam Interact. Mater. At.* **1987**, *24–25*, 763.
[51] M. L. Crespillo, J. T. Graham, Y. Zhang, W. J. Weber, *J. Lumin.* **2016**, *172*, 208.
[52] W.-K. Chu, M. James W., N. Marc-A., *Backscattering Spectrometry*, Academic Press, **1978**.
[53] M. Mayer, *MAX-PLANCK-INSTITUT FÜR PLASMAPHYSIK (GARCHING BEI MÜNCHEN)* **1997**.
[54] M. Mayer, *Nucl. Instrum. Methods Phys. Res. Sect. B Beam Interact. Mater. At.* **2014**, *332*, 176.
[55] L. C. Feldman, J. W. Mayer, S. T. Picraux, *Materials Analysis by Ion Channeling: Submicron Crystallography*, Academic Press, New York, **1982**.

**Table 1.** The RBS composition analysis on Te-rich and Te-deficient samples. Layer#1/ FTS is on top of layer#2/ $Bi_2Te_3$ or $Bi_4Te_3$.

| Sample label | Bi:Te flux ratio | Layer# and composition | Elements, Concentration percentage (respect to 1) | Actual Se $x$ content | Thickness (nm) |
|---|---|---|---|---|---|
| A | 1:5 | #1/ Fe(Se,Te) | [Fe]= 0.381, [Se]= 0.078, [Te]= 0.541 | 0.126 | 17.85 |
| | | #2/ $Bi_2Te_3$ | [Bi]= 0.432, [Te]= 0.568 | | 17.08 |
| B | 1:2 | #1/ Fe(Se,Te) | [Fe]= 0.390, [Se]= 0.072, [Te]= 0.538 | 0.118 | 18.49 |
| | | #2/ $Bi_2Te_3$ | [Bi]= 0.440, [Te]= 0.560 | | 17.81 |
| C | 1:1.5 | #1/ Fe(Se,Te) | [Fe]= 0.375, [Se]= 0.071, [Te]= 0.554 | 0.114 | 21.55 |
| | | #2/ $Bi_4Te_3$ | [Bi]= 0.635, [Te]= 0.365 | | 11.47 |

# Supporting Information for

# Interfacially enhanced superconductivity in Fe(Te,Se)/Bi$_4$Te$_3$ heterostructures

*An-Hsi Chen[1], Qiangsheng Lu[1], Eitan Hershkovitz[2], Miguel L. Crespillo[3], Alessandro R. Mazza[1,4], Tyler Smith[1], T. Zac Ward[1], Gyula Eres[1], Shornam Gandhi[2], Meer Muhtasim Mahfuz[2], Vitalii Starchenko[5], Khalid Hattar[3], Joon Sue Lee[6], Honggyu Kim[2], Robert G. Moore[1,\*], Matthew Brahlek[1,#]*

[1]Materials Science and Technology Division, Oak Ridge National Laboratory, Oak Ridge, TN, 37831, USA
[2]Department of Materials Science and Engineering, University of Florida, Gainesville, FL, USA
[3]Department Nuclear Engineering, The University of Tennessee, Knoxville, TN, 37996, USA
[4]Materials Science and Technology Division, Los Alamos National Laboratory, Los Alamos, NM, 87545, USA
[5]Chemical Sciences Division, Oak Ridge National Laboratory, Oak Ridge, TN, 37831, USA
[6]Department of Physics and Astronomy, The University of Tennessee, Knoxville, TN, 37996, USA
Correspondence should be addressed to [\*]moorerg@ornl.gov, [#]brahlekm@ornl.gov

4. Molecular beam epitaxy growth
5. Experimental Methods
    a. X-ray diffraction
    b. Scanning transmission electron microscopy
    c. Transport
    d. Angle-resolved photoemission spectroscopy
    e. Rutherford backscattering spectroscopy
6. Additional structural data

**Molecular beam epitaxy (MBE) growth**

A homebuilt MBE was used to grow the high-quality films on 5×5 mm$^2$ c-plane Al$_2$O$_3$. To understand the relation of Fe(Te,Se) to Bi$_2$Te$_3$, the film growth was optimized as follows. First, prior to growth, the fluxes were carefully calibrated using in-situ quartz crystal microbalance (QCM). QCM measurements are precise and reproducible, yet they may be systematically off in accuracy, so the fluxes were calibrated ex-situ using X-ray reflectivity and Rutherford backscattering spectroscopy (RBS). The Al$_2$O$_3$ substrates were then heated to 600 °C in a Te flux for 15 minutes in ultra-high vacuum (pressure < 1×10$^{-9}$ Torr), then the sample was cooled to temperature 145 °C where a thin, 3 quintuple layer (QL, 1 QL ≈ 1 nm), Bi$_2$Te$_3$ buffer layer was grown. The Bi$_2$Te$_3$ buffer layer was grown with a flux ratio Bi:Te = 1:2.0, with a slight Te excess above the stoichiometric ratio of 1:1.5 since this temperature was close to the point where the sticking coefficient of both Bi and Te was unity. The samples were then heated to 225 °C where 10 QL of $Bi_2Te_3$ were grown with Bi:Te ratios varying from 1:0.5 to greater than 1:20. The $Bi_2Te_3$ growth was followed by growth of Fe(Te,Se) whose thickness ranged from around 20 nm (30 unit cells (UC)) down to a single unit cell. The growth temperature of Fe(Te,Se) ranged from 200 °C to 225 °C, as discussed below. For the Fe(Te,Se) growth, a flux ratio of Fe:(Te+Se) ≈ 1:2 was used with Fe flux fixed at $1\times10^{13}$ $1/cm^2s$. After the growth was completed, the samples were cooled to room temperature then capped with a ~20 nm polycrystalline Te layer to protect from atmospheric contamination. This recipe resulted in the highest structurally and electronic quality, where both the $Bi_2Te_3$ buffer layer and Fe(Te,Se) properties can be systematically tuned.

**Experimental Methods**

  a. **X-ray diffraction**

The crystalline structure was studied using X-ray diffraction (XRD) on a Malvern Panalytical X'Pert3 with Cu-$k_{\alpha1}$ radiation ($\lambda$=0.15405 nm).

  b. **Scanning transmission electron microscopy**

Cross-section transmission electron microscopy samples were prepared using a FEI Helios Dualbeam Nanolab 600 focused ion beam system with a final milling energy of 2 keV. High-angle annular dark-field (HAADF) imaging in scanning transmission electron microscopy (STEM) was performed on a Themis Z equipped with a 5th-order probe spherical aberration corrector, operated at an acceleration voltage of 200 keV with a convergence semi-angle of 25 mrad. The positions of the atomic columns in the HAADF-STEM images were determined by two-dimensional Gaussian peak fitting using a custom Python script.

**Bond length mapping**: Figures. S1(a) and (b) are bond length maps of the Te-rich and Te-deficient samples, respectively. These higher field of view images consist of over 4000 bond length measurements, providing a more statistically robust result. To quantify the difference in extent of $Bi_2Te_3$ to $Bi_4Te_3$ reduction in the two different samples, we calculated the total standard deviation of the bonds in each row and plotted them in Figs. S1(c) and (d). The average standard deviation of each row was found to be 0.21 Å and 0.31 Å for the Te-rich and Te-deficient samples, respectively. As described in Figure 2, the reduction of $Bi_2Te_3$ to $Bi_4Te_3$ leads to a deviation in the bond structure across a row of atomic columns. A larger standard deviation value indicates more instances of the $Bi_2Te_3$ to $Bi_4Te_3$ reduction happening in the Te-deficient samples.

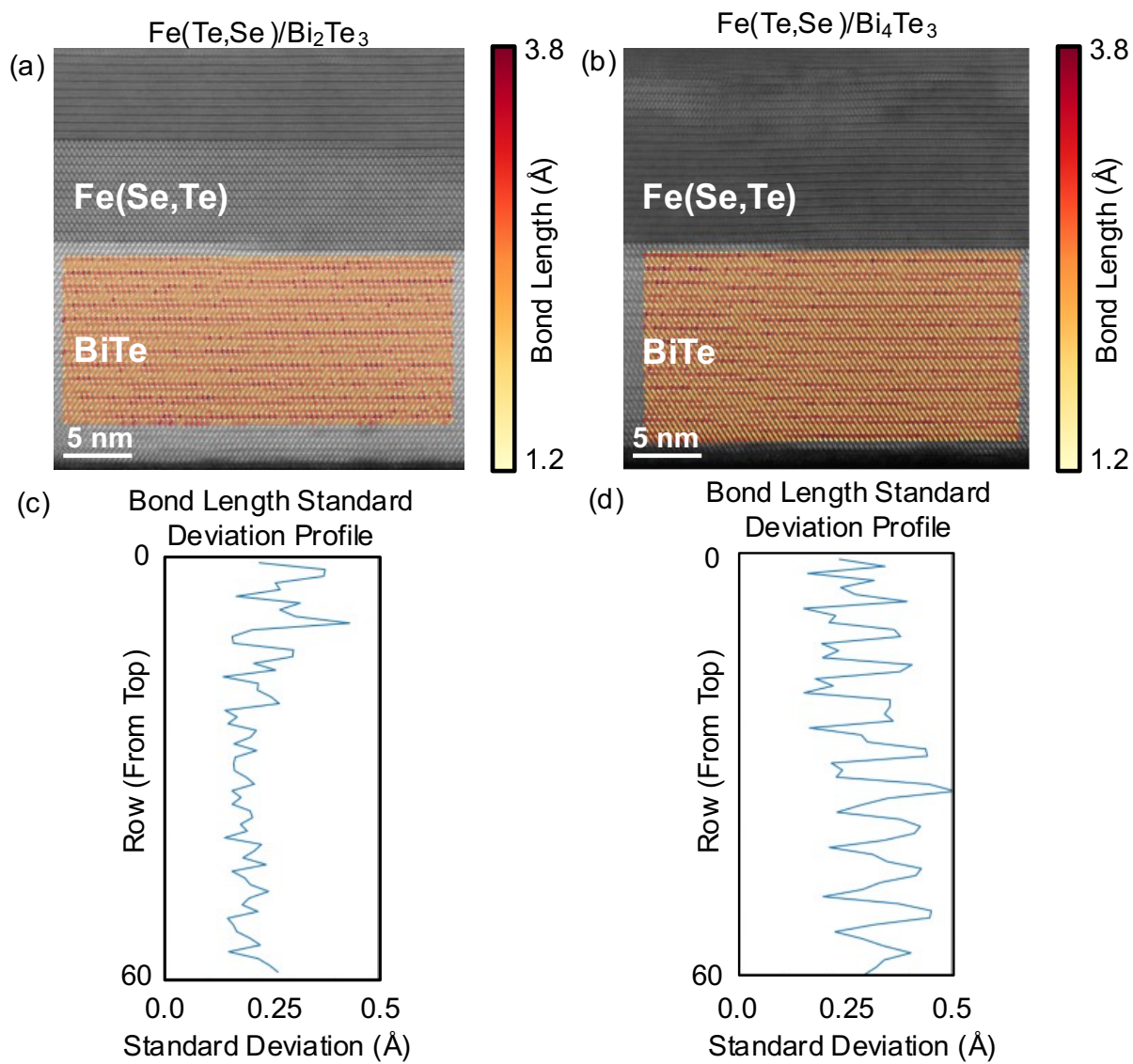

**Figure S1** HAADF-STEM images with bond length maps overlaid for both the (a) $Bi_2Te_3$ and (b) $Bi_4Te_3$. The corresponding bond length standard deviation profile, measured for each atomic row, is shown below in (c) and (d), respectively. The difference in average standard deviation between the Te rich and Te deficient samples is 0.21 to 0.31 Å, respectively, indicating a higher population of the $Bi_4Te_3$ in the Te rich sample.

c. Transport

The temperature dependent resistivity was measured using a Quantum Design Physical Properties Measurement System (PPMS) with temperature ranging from 2 K to 300 K and magnetic field up to 9 T. The electrical contacts were made with thin indium wire pressed at four corners of sample in van der Pauw method.

### d. Angle-resolved photoemission spectroscopy

The angle-resolved photoemission spectroscopy (ARPES) was performed in-situ on thin films grown with different Bi:Te ratios under ultrahigh vacuum (~$10^{-11}$ torr) conditions. The ARPES measurements were performed at T ~ 8 K using a Scienta DA30L hemisphere analyzer with a 21.2 eV Helium lamp, where the energy resolution was ~ 10 meV, and the momentum resolution was ~ 0.01 Å$^{-1}$.

### e. Rutherford backscattering spectroscopy

RBS measurements were carried out using the 3.0 MV Pelletron (model 9SDH-2) tandem electrostatic accelerator facility, manufactured by National Electrostatics Corporation[1,2], within the Tennessee Ion Beam Materials Laboratory (TIBML) located on the University of Tennessee campus[3]. The standard RBS set-up was used where the sample was placed in a multipurpose chamber connected to a beam line located at the +15° port of the switching magnet. The experimental chamber has a goniometer (Thermionics Instruments Ltd., NW, WA, USA) with 3-axes of rotation (polar, azimuthal and flip rotation axes, allowing the sample to be positioned at the proper angle (crystallographic orientation). The RBS spectra measurements were carried out with a collimated 4.5 MeV He$^+$ beam with the particle flux of 6.9×10$^{12}$ cm$^{-2}$ s$^{-1}$ extracted from the accelerator (divergence of ≈0.1°) with the probe size on the sample surface of 1.0×1.0 mm$^2$. Each spectrum was collected accumulating a charge of 20 μC. A fixed silicon surface barrier detector (ORTEC PIPS, USA) with an energy resolution of 15 keV placed in IBM geometry[4] and positioned at scattering angle of 155° relative to the incoming beam with a solid angle of 1.1×10$^{−3}$ sr was used to collect the signal of the backscattered He$^+$ yield from the samples.

The resulting spectra were fitted by SIMNRA software v7.02[5,6] with calculations using the resonant cross section provided by SigmaCalc 2.0. RBS analysis has been performed to characterize the atomic concentration and compositional depth profile of Bi, Te, Se, and Fe elements within the layers of the multi-layer (MLs) systems. The backscattered yield, proportional to the differential cross-section of the elements, is also proportional to their atomic mass (M) and the elemental concentration present in the layer [5,6]. The backscattered energy is proportional to the atomic mass (*M*) and is assigned to a corresponding energy channel in the measured spectrum[4,7]. Thus, to maximize the atomic mass resolution (separation) for elements with high $M$, a high energy 4.5 MeV He$^+$ ion beam was used.

The Se/Te ratio was extracted directly through the simulation. The value quoted in the text was the Se/(Te+Se) ratio taken within the Fe(Te,Se) layer. Here, the Se/(Te+Se) ratio is likely a slight underestimate of the realistic concentration. Specifically, the simulation likely overestimates the Te concentration in the Fe(Te,Se), which stems from the large concentration of Te in samples, including the Bi$_2$Te$_3$ underlayer and the Te capping layer. We also considered the direct Se/Fe ratio, which for an ideal stoichiometric sample should also give the relative concentration. Thermodynamically, FeTe tends to have iron-rich stoichiometry (Fe$_{1.1}$Te) in bulk. Thus, the Se/Fe ratio can range from 0.18 (iron-rich, for cation to anion ratio of 1.1:1) to 0.20 (1:1) Hence, this is systematically inaccurate and may be an overestimate. Altogether, as mentioned in the main text, we estimate that the error in stoichiometry is around ±5%, which is well within the accuracy of previous reports (many reports on thin films neglect direct measurements of stoichiometry) as well as bulk crystals.

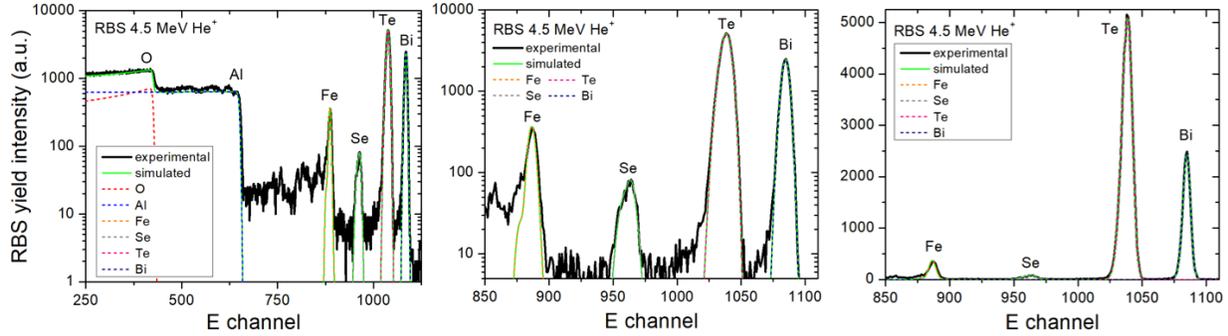

**Figure S2**. Sample RBS data and corresponding simulations for sample A in Table 1 of the main text. Left, full spectrum. Center, zoom-in about the peaks that correspond to the elements in the film. Right, same data but in linear scale.

**Additional structural data**

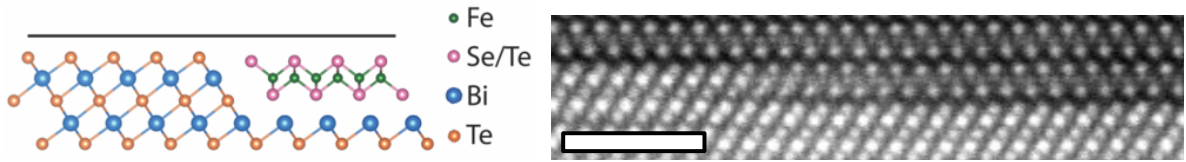

**Figure S3**. High magnification HAADF-STEM image of a select feature of the interface of $Bi_2Te_3$ sample shown in Figure S2 where a portion of the terminal monolayer of $Bi_2Te_3$ QL structure splits and intergrows with the first monolayer of FTS. The cartoon at the left was reproduced from G. Chen et al. npj Quantum Materials 7, 110 (2022). The scalebar on the image represents 2 nm.

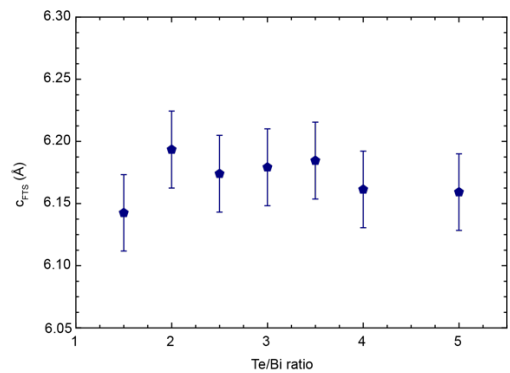

**Figure S4.** c-axis lattice parameter of Fe(Te,Se) as the function of flux Te/Bi ratio used for the growth of the $Bi_2Te_3$.

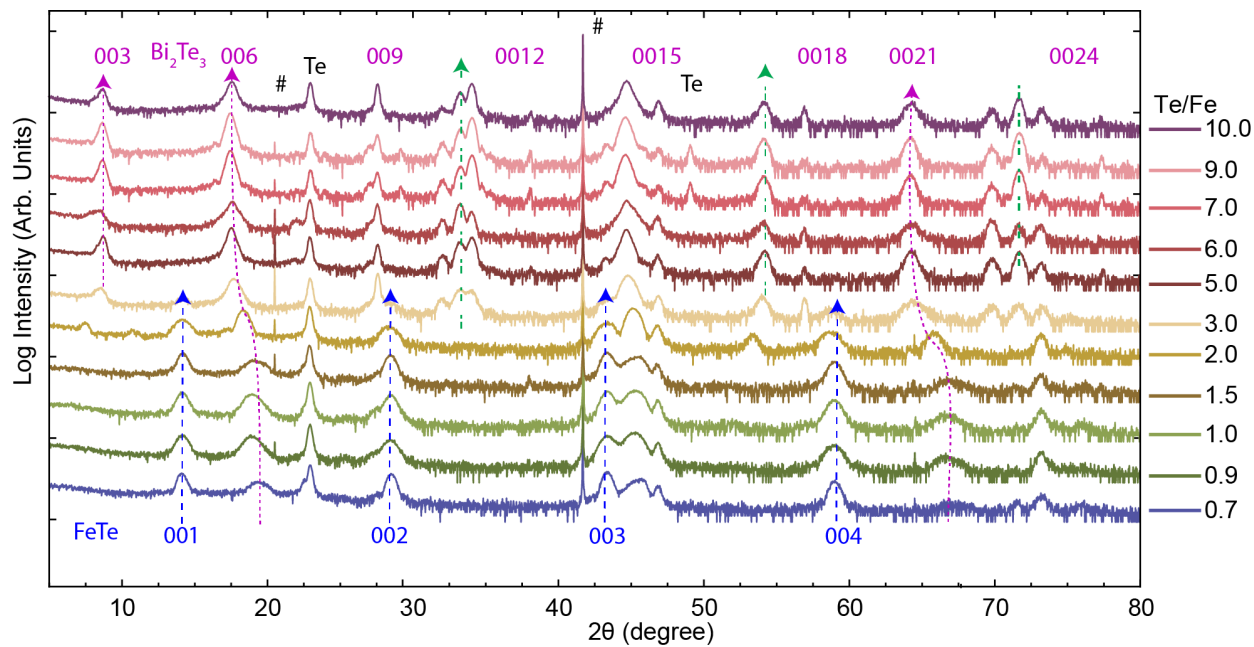

**Figure S5.** The $2\theta\text{-}\theta$ data for FeTe/Bi$_2$Te$_3$ growth with different Te/Fe ratios of FeTe layer. The dashed pink arrows highlight the shift of {003} Bi$_2$Te$_3$, blue arrows indicate the position of {001} FeTe up to a flux ratio of around 3 and green arrows highlight the appearance of Fe-Te with excess Te. The hash # symbol mark out the diffraction from c-plane Al$_2$O$_3$ substrate.

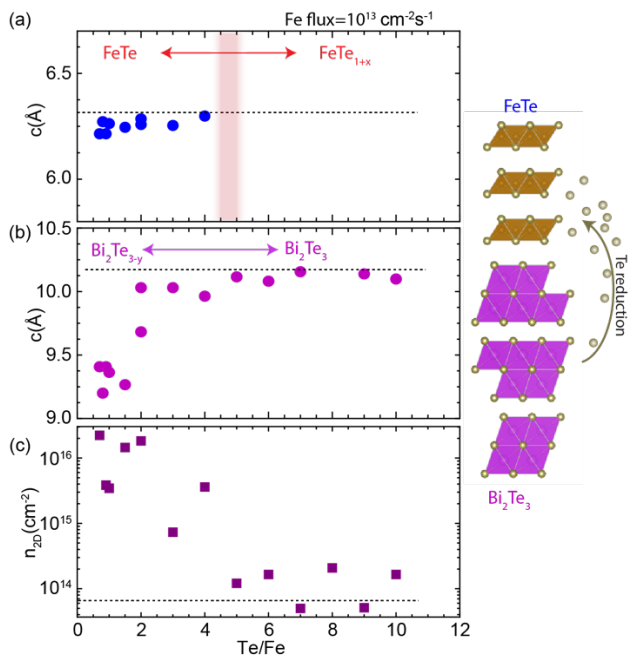

**Figure S6.** The crystalline and electronic properties of FeTe/$Bi_2Te_3$. (a) The c-plane distance of FeTe and (b) $Bi_2Te_3$ versus Te/Fe ratio is calculated from $2\theta$ value of X-ray diffraction in Figure S4 assuming the $Bi_2Te_3$ structure. (c) The carrier density versus Te/Fe ratio is estimated from Hall effect measurements. The Te-poor condition has a higher carrier density over Te-rich. (d) The schematic illustrates the Te diffusion from $Bi_2Te_3$ to Te-poor FeTe layer.

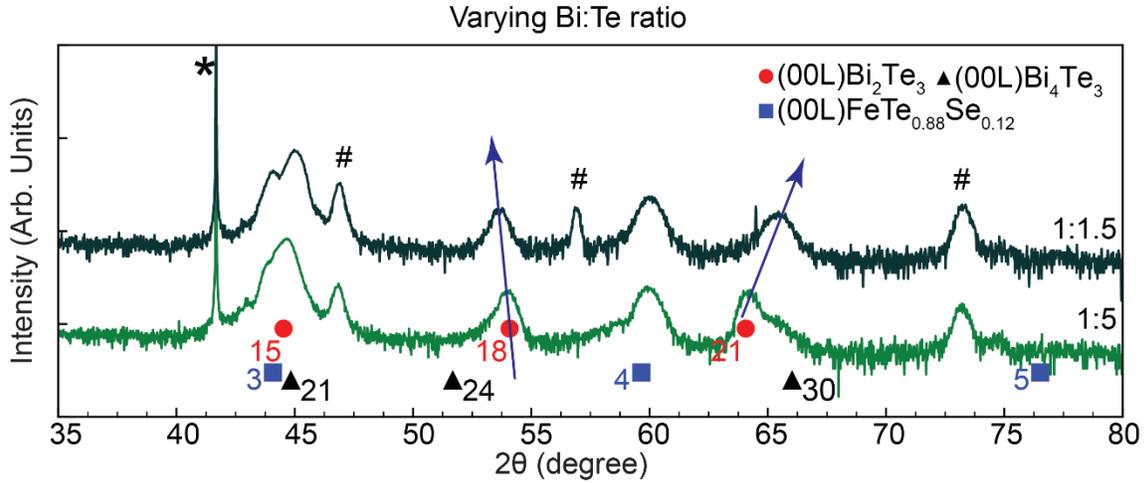

**Figure S7.** The $2\theta$-$\theta$ data for Fe(Te,Se)/$Bi_2Te_3$ with $x$ fixed at 0.12 versus Bi:Te flux ratio at higher $2\theta$ range. The blue squares mark {001} FTS reflections, red circles mark {003} $Bi_2Te_3$ reflections, and the black triangles mark {003} $Bi_4Te_3$ reflections. The # symbol indicates Te reflections of the capping layer and the * symbol marks the $Al_2O_3$ substrate reflections. The blue arrows highlight the peak shifts that occur.

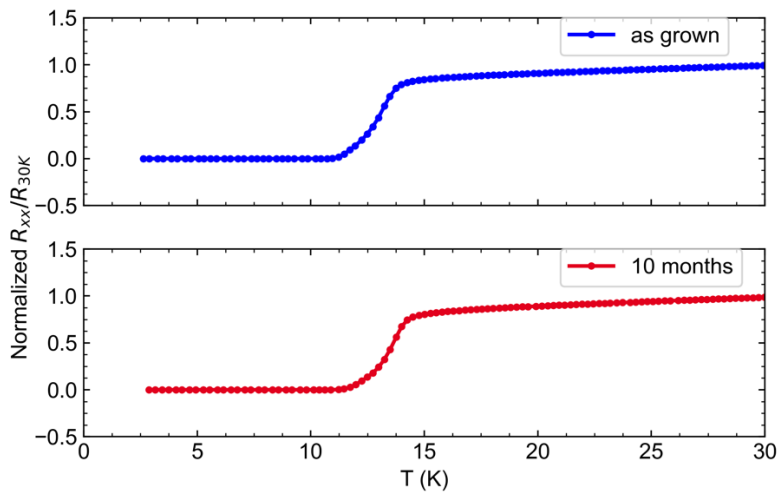

Figure S8. Normalized resistance versus temperature measurement of Fe(Te,Se)/$Bi_2Te_3$ with 25% Se. The top row is measured directly after growth and the bottom row is measured after 10 months. The normalized resistance is defined as resistance ($R_{xx}$) divided by the resistance at 30K ($R_{30K}$).

**References**

[1] G. A. Norton, R. E. Daniel, R. L. Loger, J. B. Schroeder, *Nucl. Instrum. Methods Phys. Res. Sect. B Beam Interact. Mater. At.* **1989**, *37–38*, 403.
[2] J. B. Schroeder, C. W. Howell, G. A. Norton, *Nucl. Instrum. Methods Phys. Res. Sect. B Beam Interact. Mater. At.* **1987**, *24–25*, 763.
[3] M. L. Crespillo, J. T. Graham, Y. Zhang, W. J. Weber, *J. Lumin.* **2016**, *172*, 208.
[4] W.-K. Chu, M. James W., N. Marc-A., *Backscattering Spectrometry*, Academic Press, **1978**.
[5] M. Mayer, *MAX-PLANCK-INSTITUT FÜR PLASMAPHYSIK (GARCHING BEI MÜNCHEN)* **1997**.
[6] M. Mayer, *Nucl. Instrum. Methods Phys. Res. Sect. B Beam Interact. Mater. At.* **2014**, *332*, 176.
[7] L. C. Feldman, J. W. Mayer, S. T. Picraux, *Materials Analysis by Ion Channeling: Submicron Crystallography*, Academic Press, New York, **1982**.